\documentstyle[aps,epsf,twocolumn,prl]{revtex}
\draft

\newcommand{\bb}{\begin{eqnarray}}
\newcommand{\ee}{\end{eqnarray}}
\newcommand{\p}{\partial}
\newcommand{\eps}{\epsilon}
\newcommand{\vv}{\mu^*}
\newcommand{\dd}{\nu}
\newcommand{\MU}{\beta\Delta\mu}

\newcounter{subequation}

\newcommand{\startnumbering}{ \setcounter{subequation}{1}
\renewcommand{\theequation}{\arabic{equation}-\alph{subequation}} }
\newcommand{\finishnumbering}{ \renewcommand{\theequation}{\arabic{equation}}
\setcounter{equation}{\value{equation}} }

\begin{document}
\title{Polymer Translocation through a Pore in Membrane}
\author{W. Sung and P. J. Park}

\address{Department of Physics, Pohang University of Science and
Technology, Pohang 790-784, Korea}

\maketitle
\begin{abstract}
\indent We construct a new statistical physical model of
polymer translocation through pore
in membrane
treated as the diffusion process
across a free energy barrier.
We determine the translocation time in terms of chain flexibility
yielding an entropic barrier, as well as in terms of the driving 
mechanisms such as 
transmembrane chemical potential difference and 
Brownian ratchets.
It turns out that, while 
the chemical potential differences induce pronounced effects 
on translocation due to the long-chain nature of the polymer,
the ratchets suppress this effect and chain flexibility.

\vspace{0.2cm}
\noindent PACS number(s): 05.40.+j, 83.10.Nn, 87.22.-q, 87.22.Fy\\
(To appear in Phys. Rev. Lett.)\\
\end{abstract}
The process of polymer translocation into or across biomembranes
is a problem of considerable importance to 
a multitude of biological functions.
Proteins are transported across 
a cellular membrane and endoplasmic reticulum, 
while RNAs across a nuclear membrane
after their synthesis\cite{Singers90,Schatz}. There are 
similar macromolecular transport mechanisms
in drug delivery, as well as in biotechnology of gene transfer\cite{Chang}
where it is fundamental to understand how DNAs can be incorporated 
into cells.
It is a highly complex process with specificity
involving conformational changes of the translocating polymers that
can occur in both cis and trans sides as well as inside of the membrane.

Although the translocation apparatus have been suggested and examined
empirically in a great
variety\cite{BIOPAPER}, only recently there
have been a few efforts to investigate quantitatively the driving force of
translocation on physical grounds\cite{BS95,SPO92}.
Baumg\"{a}rtner and Skolnick\cite{BS95} studied via simulation the
translocation of polymer directly through lipid bilayer,
driven by the
concentration imbalance of lipids that exists at high-curvature regions 
in membrane.  On the other hand,
Simon, Peskin and Oster(SPO)\cite{SPO92} 
considered protein translocation through a translocation channel or pore, and 
postulated that its driving force 
is random thermal motion rectified by 
'ratchets' which give rise to directional diffusion.
The origin of this so called Brownian ratchets(BRs) is a chemical 
asymmetry, i.e., 
if specific predetermined segments of the protein
cross the membrane, chemicals such as 
chaperones bind on the segments to prevent their 
backward diffusion to the cis side of the membrane. 

SPO considered rigid proteins, leaving out the effects of three
dimensional chain conformations and the associated 
flexibility and entropy. In this Letter we incorporate these important
effects by considering the flexible-polymer model. 
Along with the BR mechanism, we also incorporate the more ubiquitous
kind of asymmetry due to
transmembrane chemical potential difference, which naturally
exists in biomembranes, due to e.g., electrochemical gradients, membrane
potentials and protein conformational changes.
We aim at an analytic, quantitative theory on the basis of statistical
physics of polymer and stochastic processes.
To highlight these flexibility and asymmetry effects on translocation, 
we consider a simple, but tenable model for the membrane: 
a rigid wall of negligible thickness 
with a pore, which is assumed to be
small enough to allow only a single segment passage.
The interaction between membrane and polymer segments 
is considered to be only of steric origin, i.e., the segments
cannot cross membrane except through the pore. 
We describe the translocation dynamics  
as a stochastic process crossing
the free energy barrier 
calculated from chain configuration partition function.
The translocation time, given as the mean first passage time for this
barrier crossing, is obtained from the Fokker-Planck equation 
that we formulate below. 
The initial targeting of nascent chain 
to the pore is regarded as a separate process and is excluded in this study. 
The controversies\cite{Schatz,Glick} 
over chain conformations and chaperone functions
go beyond this investigation, which is mainly concerned with
finding some nonspecific physical principles behind translocation.

{\it Free Energy Barrier of Polymer Translocation} ---
The conformation of a flexible polymer during its translocation 
is significantly affected by steric interaction with the membrane,
leading to a reduction of the polymer entropy 
and increase of its free energy. 
We adopt, as our model,
an ideal chain with $N( \gg 1)$ Kuhn segments each with length $b$.
First consider a chain with $n$ Kuhn segments with the initial segment
anchored on  a rigid wall introduced in $yz$ plane.
With the boundary condition(BC) that the other segments do not cross the
surface, the $G({\bf r},{\bf r}_0; n)$, 
the probability of finding the end segment at 
${\bf r}$, given initial one at ${\bf r}_0$ on surface,
is obtained using the image method\cite{Mathew};
it is given as the probability for all configurations in free space,
the Gaussian distribution $ G_0( {\bf r}, {\bf r}_0; n ) = (2\pi n b^2/3)^{-3/2}
\exp[- 3({\bf r}-{\bf r}_0)^2/(2 n b^2)]$,
minus the probability for the chain crossing the surface 
$G_0( {\bf r}, -{\bf r}_0; n )$,
\bb
G({\bf r},{\bf r}_0; n) &=&  
G_0( {\bf r}, {\bf r}_0; n ) - G_0( {\bf r}, -{\bf r}_0; n ) 
\nonumber \\ 
&=& \left[\frac{2\pi nb^2}{3}\right]^{-3/2} \frac{6x\eps}{nb^2} 
\exp(-\frac{3 {\bf r}^2}{2nb^2}),
\ee
where ${\bf r}_0 = (\eps,0,0)$, 
$\eps$ is an arbitrarily small distance 
of the anchored segment from the surface. 
The steric constraint factor of chain, given as
$ Z_S(n) = \int_{x>0} G({\bf r},{\bf r}_0; n) d{\bf r}<1$,
scales as $n^{-1/2}$. 
In the absence of the constraint the partition function
is given by $Z_B(n) \sim \exp(-\beta n \mu)$, where $\beta = 1/k_B T$ and
$\mu$ is the chemical potential per segment defined by  
$\mu = (\p F(n)/\p n)_T$ in the limit $n\rightarrow \infty$.
The $F(n)$ is the free energy given from the full partition function, 
$ F(n) = -k_BT \log \left[Z_S(n) Z_B(n) \right] 
     = {1 \over 2}k_BT \log n + \mu n + \mbox{const}, $
where the constant term is independent of $n$.

\begin{figure}[b]
\vspace{6cm}
\includegraphics{fig/fig1.ps} 
\caption{Schematic figure of the configuration of a 
   translocating polymer.}
\end{figure}
The whole chain during translocation can be decomposed 
into two independent end-anchored chains each 
in the opposite half spaces.
For the decomposition into $n$ and $N-n$ segments as shown in Fig. 1,
the total free energy is
\bb
{\cal F}(n) &=& F(n) + F(N-n)\\
	&=& {1\over 2} k_BT \log \left[n(N-n)\right] + n \Delta \mu + \mbox{const}, 
\label{F}
\ee
where $\Delta \mu$ is the excess chemical potential 
per segment of trans side relative to that of cis side.
The free energy with $\Delta \mu = 0$ has a
symmetric barrier of entropic origin which, for a long chain,
is nearly flat except near  $n=1$ or $n=N-1$(B of Fig. 2). 
As also shown in Fig. 2,~~for
\begin{figure}[b]
\vspace{5.2cm}
\includegraphics{plot/f/f.ps}
\caption{Free energy ${\cal F}(n)$ as a function of 
         translocation coordinate $n$. ($ N=1026$,
          A: $\beta \Delta\mu=10/N$, B: $\beta \Delta\mu=0$, 
	  C: $\beta \Delta\mu=-10/N$ )}
\end{figure}
\noindent a very long chain, a minute chemical
potential difference(e.g.,~$\Delta \mu=10^{-2}k_BT$) 
can break the barrier shape symmetry
and its contribution can dominate the free energy.
This contribution, which does not not appear for a polymer
in homogeneous media, can yield
pronounced effects on a translocating polymer we shall see below.

 
{\it Stochastic Model for Translocation Dynamics} ---
For the long-time scale behavior of translocation,
we construct a coarse-grained description in terms of 
the translocated segment number(translocation coordinate) 
$n$ adopted as a relevant stochastic
variable and in terms of the associated free energy barrier.
It can be treated as a diffusive
random process, which is described by
a Fokker-Planck equation for $P(n,t)$, the probability distribution
of $n$,
\bb
{\p\over\p t} P(n,t) = {\cal L}_{FP}(n) P(n,t),
\ee
where ${\cal L}_{FP}(n)$ 
is the operator, $ {\cal L}_{FP}(n) = 
1/b^2 (\p/\p n) D(n) \exp(-\beta {\cal F}(n))
(\p/\p n) \exp(\beta {\cal F}(n))$.
Here, $D(n)$ is the chain diffusivity during translocation.
In the case that the $D$ remains constant, 
it is given by $D = k_BT/\Gamma \sim N^{-\dd}$,
where $\Gamma$ is the chain friction
proportional to $N^{\dd}$. 
The exponent $\dd$ is $1$ if
the hydrodynamic interaction between the segments is neglected (as in the
Rouse model), 
and is $1/2$ if it is included (as in the Zimm model)\cite{DoiEdward}.

The mean first passage time $\tau (n, n_0)$, which is defined as
the time for diffusion from the coordinate $n_0$ to $n$, is
obtained by solving the equation\cite{Risken84}
${\cal L}^{\dag}_{FP}(n_0) \tau (n, n_0) = -1$,
where
${\cal L}^{\dag}_{FP}(n_0) = 1/b^2 \exp(\beta {\cal F}(n_0))
(\p / \p n_0) D(n_0) \exp(-\beta {\cal F}(n_0)) (\p / \p n_0)$.
To obtain the translocation time for the case that
only the front segment in
trans side is ratcheted,
we assign the reflecting and absorbing BCs
respectively at $n=1$ and $n=N-1$:
$ J( n=1, t ) =-\left[ (D(n)/b) \left( \p /\p n+ \beta \p {\cal F}/ \p n\right) P(n,t) \right]_{n=1} = 0$, and 
$P( n=N-1, t) = 0$.
Under these BCs, the translocation time,
defined by $\tau \equiv \tau(N-1,1)$ is integrated to be
\bb
\tau &=& b^2 \int_1^{N-1} dn \frac{1}{D(n)} e^{ \beta {\cal F}(n) } 
\int_1^{n} dn' e^{ -\beta {\cal F}(n') }.  
\label{MFPT}
\ee

Let us first assume, for simplicity, that $D$ does not change in
the course of translocation.
In case of rigid chain without chemical potential difference
$\Delta\mu$, 
${\cal F}(n)=\mbox{const}$, the translocation 
time is simply reduced to 
$ \tau = L^2/2D \sim L^{2+\dd}$,
the result for the one dimensional diffusion of a single
Brownian particle.
Here $L=Nb$ is length of the whole chain.
To incorporate the chain flexibility effect, 
the free energy function in Eq. \ref{F} should be included in 
Eq. \ref{MFPT}, resulting in, for $\Delta\mu=0$,
\bb
\tau(\Delta \mu=0) = {\pi^2 \over 8} \frac{L^2}{2D} \sim L^{2+\dd}.
\ee
While the length scaling behavior of the translocation time of 
flexible chain is same as
that of rigid chain,
the prefactor of $\pi^2/8$ indicates that the chain flexibility
retards translocation by $23\%$.
This trend is opposite to what SPO obtained, 
due to the entropy effect associated with 
the three dimensional chain conformation
which they did not include\cite{SPO92}.
The translocation time is proportional to 
$N^{2+\dd}$ and, remarkably, with $\dd=1$ this
scaling behavior is identical to that of chain reptation time in
entangled polymer systems\cite{deGennes,DoiEdward}.

If there is a nonvanishing chemical potential difference,
the translocation time can be calculated, having
the analytical expressions for limiting cases,
\begin{flushright}
\startnumbering
\bb
\hspace{0.5cm}
\tau(\vv) = \left\{ 
\begin{array}{cc}
		\frac{\pi^2}{8} \frac{L^2}{2D} \left( 1+{32\over 9\pi^2} \vv \right) &, |\vv| \ll 1  \\
		\frac{L^2}{2D} \frac{2}{|\vv|} &, \vv \ll -1 \\
		\frac{L^2}{2D} \frac{2}{{\vv}^2} \exp(\vv) &,  \vv \gg 1,
\end{array} \right. 
\begin{array}{cc}
\mbox{ \hspace{0.5cm} (\theequation)}     \label{Eq:limit1} \\ 
\stepcounter{subequation}
\mbox{ \hspace{0.5cm} (\theequation)}     \label{Eq:limit2} \\
\stepcounter{subequation}
\mbox{ \hspace{0.5cm} (\theequation)}     \label{Eq:limit3} 
\end{array} 
\nonumber
\ee
\finishnumbering
\stepcounter{equation}
\end{flushright}
\vspace{-0.5cm}
\noindent where $\vv \equiv N\beta\Delta\mu$.
When the chemical potential per segment is reduced
on trans side, the translocation time as given by Eq. \ref{Eq:limit1}
and Eq. \ref{Eq:limit2}
encounters a
crossover in the scaling behavior from 
$\tau \sim N^{2+\dd} \sim L^{2+\dd}$
to $\tau \sim N^{1 + \dd} \sim L^{1 + \dd}$. 
As shown in Fig. 3, this crossover occurs
around $\vv = 1$
corresponding  to $\Delta \mu = k_B T / N$,
a very minute chemical potential difference for a long chain.
This remarkable sensitivity of translocation to chemical potential
asymmetry is even enhanced
for the opposite case of higher chemical potential on 
trans side.
Consider, for an example, a chain with $N = 10^3$, and 
$|\Delta \mu | = 10^{-2} k_B T$, then $|\vv| = 10$.
While, this small chemical potential difference with negative sign, speeds
up the polymer translocation by the factor $\tau(\vv=-10)/\tau(\vv=0) = 0.18$, 
the one with positive sign slows it down
by the factor of $1191$.
Regardless of chain flexibility, this extreme sensitivity, 
already implied in Fig. 2, is a cooperative phenomenon
arising from chain connectivity; 
the segments respond all hand in hand 
(involving the scaling variable $\mu^{*}=N\beta \Delta \mu$ in Eq.~7) 
rather than as individuals (involving $\beta \Delta \mu$) 
~~to~~~a~~~driving~~~asymmetry.~~~~This~~~is 
\begin{figure}[b]
\vspace{4.5cm}
\includegraphics{plot/N/n.ps}
\caption{Translocation time (in units of $b^2/2D_0$, $D_0=ND$)
versus chain length $N$ for $\dd = 1$. 
( A: $\MU =0.$, B: $\MU = -10^{-4}$, C: $\MU =
-10^{-3}$, D: $\MU = -10^{-2}$, E: $\MU = -10^{-1}$, F: $\MU =-1.0$, G:
$\MU =-10.$ ) The crossover behavior from $\tau \sim N^3$ to $\tau
\sim N^2$ occurs when $N$ is around $k_BT/ |\Delta\mu |$.  }
\end{figure}
\noindent reminiscent of the cooperative effect of a slight segmental bias 
that gives rise to fast protein folding as proposed by Zwanzig,
Szabo and Bagchi\cite{Zwanzig}.

The chain diffusivity can also change
during translocation. Adopting the Rouse model, 
$D(n)^{-1}=N^{-1} (n D^{-1}_t +(N-n)D^{-1}_c)$ where $D_c$ and 
$D_t$ are the diffusivities of the whole chain in cis and trans sides
respectively. The effect of the change, $\Delta D = D_t-D_c$, on $\tau$ can be incorporated analytically,
but the result does not affect the dramatic effect of $\Delta\mu$ 
discussed above.
The relative insensitivity of $\tau$ to $\Delta D$ is obvious since,
while $\Delta\mu$ appears exponentially, $D$ is involved inversely
in Eq. \ref{MFPT}.

{\it Many-Ratchet Effect} ---
The BR mechanism, which was originally suggested by SPO as a nonspecific
driving mechanism for biased diffusion, 
assumes fast chemical
binding of chaperones on chain entering the trans side of membrane\cite{SPO92}.
The binding sites are assumed to be 
uniformly distributed with an interval of $\delta$ along the chain.
To incorporate this ratchet mechanism within our model for
the case of instantaneous action of ratchets without dissociation, 
the whole space of translocation coordinate is divided
into intervals  of length $\delta = L/M$, where $M$ is the number 
of binding sites. 
Then the range of $i$-th interval is $(i-1)\alpha +1< n < i \alpha +1$, 
where $i = 1, 2, 3, \cdots, M$, and $\alpha = N/M$ is the number of
polymer segments in each interval.
The dynamics is now consecutive translocation(uni-directional diffusion)
of each interval subject to the free energy therein, 
as well as to the BCs at both borders of the interval,
reflecting BC at the left border and the absorbing BC at the right.
These BCs are written as, 
$ J(n=(i-1)\alpha+1,t) = 0$ and $ P(n=i \alpha+1,t) = 0$,
for all intervals.

Assuming $D$ is constant, the translocation time of the whole polymer 
is then $ \tau = \sum_{i=1}^{M} \tau_i $,
where 
\bb
\tau_i = \frac{b^2}{D} \int_{(i-1)\alpha+1}^{i\alpha+1} dn e^{\beta {\cal F}(n)}
\int_{(i-1)\alpha+1}^{n} dn' e^{-\beta {\cal F}(n')}.
\ee
\begin{figure}[h]
\vspace{5cm}
\includegraphics{plot/mu/o3.ps}
\caption{$\Omega(\vv,M)$ as a function of $M$
    for different values of $\vv \equiv N\beta \Delta \mu$.
    ( $N=1026$, and A: $\vv=100.$, B: $\vv=10.74$, C: $\vv=0.$, D:
$\vv=-10.74$, E: $\vv=-100.$ ) When $M$ is sufficiently large,
    $\Omega(\vv,M)$ goes to $1$, the ratchet limit.}
\end{figure}
\noindent If the chain flexibility and $\Delta\mu$ are neglected,
i.e. ${\cal F}(n)=\mbox{const}$, 
it is reduced to $\tau = L \delta / 2D = L^2/2DM$ naturally,
the reduction by the factor of $1/M$ compared with $\tau(M=1)$,
as given by SPO.
In general the translocation time can be written as
\bb
\tau = \frac{L^2}{2DM}\Omega(\vv,M).
\ee
Numerical integration for translocation time gives
$\Omega(\vv,M)$ as depicted in Fig. 4, which clearly indicates that
the ratchets suppress the chain flexibility, 
as well as the chemical potential difference
{\it regardless of its sign}.
Most striking is the approach of $\tau$ to that 
of rigid chain($\Omega(\vv,M)=1$), i.e.,
solely the ratchet result, when $M$ is very large,
even with large negative value of $\vv$;
it runs counter to the intuition, according to which,
the negative chemical potential difference
and the ratchet mechanism add up in
series in speeding translocation.

This overriding effect of many ratchets can be better
understood by considering the Langevin equation equivalent to
the Fokker-Planck equation description,
\bb
b \Gamma \dot{n} = - \frac{1}{b} { \p {\cal F}_R ( n ) \over \p n } + \xi (t) 
\ee
where $ \xi (t) $ is the Gaussian, white noise connected to
$\Gamma$ via fluctuation-dissipation theorem(FDT),
$\left< \xi(t)\xi(0)\right> = 2 \Gamma k_BT \delta(t)$.
Confining ourselves to the case in which the chain is rigid,
the $ {\cal F}_R ( n )$ shown in Fig. 5 is the free energy(ratchet potential)
which effectively includes the BCs 
due to ratchets.  For the reflecting and absorbing BCs we considered,
the step height $h$ is infinity. (But for general consideration
and more realistic ratchet activities,
it can be put to be finite. 
The similarity of this ratchet potential to
those employed for ratchet-driven motor 
proteins\cite{Astumian} is~~remarkable.)~~The~~FDT~~assures~~the~~approach~~to 
\begin{figure}[h]
\vspace{6.5cm}
\includegraphics{fig/ratchet.ps}
\caption{The free energy barrier of a rigid chain with $M =5$ ratchets
and with chemical potential difference $\Delta \mu$. Here
$\alpha = N/M$ is the number of chain segments per ratchet and $\alpha \Delta \mu$,
$h(\gg \alpha \Delta \mu)$ are the barrier heights due to
asymmetries arising from chemical potential difference and from ratchet
activity, respectively.}
\end{figure} 
\noindent 
equilibrium, that is, under the potential ${\cal F}_R ( n )$ 
of Fig. 5, $n$ undergoes the rectified diffusion to the right, 
regardless of $\Delta \mu$.
When $M$ is very large, so that $\alpha \Delta \mu = k_B T \vv / M$,
the barrier height due to $\Delta \mu$, is very small, 
the global translocation dynamics become independent of the
details of the local potential barriers, 
yielding the $\tau$ of many ratchets, 
$\tau = L^2 / 2 D M $, or $\Omega (\vv, M) = 1$.
A calculation shows that this result tends to be valid for 
finite values of $h$ larger than $k_BT$.
 
To summarize, we have investigated mechanisms 
affecting polymer translocation
through a pore in membrane. 
It is found that, while chain flexibility, due to an entropic barrier it gives,
does retard translocation, the ratchets speed it up,
tending to reduce the flexibility and chemical potential effects
to rigid chain behavior. The transmembrane chemical potential
asymmetry, only with minute magnitude, is found to modulate dramatic
changes in translocation behaviors of long polymers,
which is a cooperative behavior arising from their chain connectivity.

The present studies were supported by the BRSI
Program, Ministry of Education, and POSTECH BRSI/Special Fund.

\end{document}